\documentclass[11pt,epsf]{JHEP3}

\def\beq{\begin{eqnarray}}
\def\eeq{\end{eqnarray}}
\def\beq{\begin{eqnarray}}
\def\ee{\end{eqnarray}}
\def\be{\begin{eqnarray}}
\def\eeq{\end{eqnarray}}
\def\bea{\begin{eqnarray*}}
\def\eea{\end{eqnarray*}}
\def\beq{\begin{equation}}
\def\eeq{\end{equation}}
\def\ba{\begin{eqnarray}}
\def\ea{\end{eqnarray}}

\title{The Intermediate Scale Branch of the Landscape}
\author{M. Dine\\ 
Santa Cruz Institute for Particle Physics, Santa Cruz CA 95064 \\ 
Email: \email{ dine@scipp.ucsc.edu} }

\abstract{Three branches of the string theory landscape have plausibly been identified.
One of these branches is expected to exhibit
a roughly logarithmic distribution of supersymmetry breaking scales.  The original
KKLT models are in this class.  We argue
that certain features of the KKLT model are generic to this branch, and that
the resulting phenomenology depends on a small set of discrete choices.  As in the
MSSM, the weak scale in these theories is tuned; a possible explanation
is selection for the dark matter density.}

\begin{document}

\section{Introduction:  Supersymmetry Breaking in the Landscape}

Implicit in traditional questions of naturalness is the notion that there is a distribution
of possible theories of
elementary particles.  The possible existence, in string theory, of an exponentially
large number of stable and metastable vacua without supersymmetry or with $N=1$ supersymmetry
in four dimensions, the ``landscape"\cite{landscape1,landscape2,landscape3,landscape4},
provides a realization of this idea.  Much is already
known about the statistics of these states\cite{douglasfirst,
Denef:2004ze,Denef:2004cf,dgt,dos}, and it is possible to assess
the tunings required
to understand the values of parameters such as the cosmological constant and the weak
scale.  It is even possible to make some cautious statements
about correlations which might lead to
experimental predictions.  There are many uncertainties, and any discussion of the landscape must be viewed
as tentative.  For example, recently, an analysis
has appeared that suggests there might be
infinite numbers of four dimensional stable or metastable states in string theory,
with all moduli
fixed and with no more than four
supercharges.  These are even claimed to be accessible to weak coupling
methods\cite{infinite}.  If correct, and if these
states are physically relevant, many of our notions of
vacuum statistics are, at best, naive.  We will comment
on these issues briefly in our conclusions, but the discussion of this paper 
will be predicated on the assumption that the number of relevant states in the landscape
is finite and conventional statistical ideas can be applied.

Three branches of the landscape have been clearly identified\cite{dgt}.  They are distinguished by
their distributions of supersymmetry breaking scales.  On one branch,
which we will refer to as
the ``broken supersymmetric branch", the bulk of the states have supersymmetry broken at very
high energies.  One can make reliable statements on this
branch only for small supersymmetry
breaking scale, $m_{3/2} \ll M_p$.  In this
regime the distribution behaves, for small cosmological constant,
$\Lambda < \Lambda_o$, as\cite{Denef:2004cf,dos}
\beq
\int dm_{3/2} P(m_{3/2}) = \Lambda_o \int dm_{3/2}^2 m_{3/2}^{10}.
\eeq
where we use units where the Planck scale is set equal to one.
On the second branch, the
``intermediate scale branch", the distribution of supersymmetry-breaking scales
is expected to be roughly logarithmic\cite{dgt}:
\beq
\int P(m_{3/2})dm_{3/2}=\int{dm_{3/2}^2 \over m_{3/2}^2} \ln(m_{3/2})
\eeq
On the third branch, the ``low scale
branch," the scale of supersymmetry breaking tends to be very small\cite{dgt},
\beq
\int P(m_{3/2})dm_{3/2}=\int{dm_{3/2}^2 \over m_{3/2}^4}
\eeq
In tree level analyses, e.g. of IIB theories on orientifolds of Calabi-Yau spaces,
the first branch corresponds to stationary points of the action with broken supersymmetry;
the second branch to points with unbroken supersymmetry and negative cosmological
constant (non-vanishing superpotential, $W$); the third branch to
states with unbroken supersymmetry and vanishing cosmological constant at tree level.
Since one expects that $N=1$ supersymmetry is often dynamically broken, these
classical distinctions
are not sharp.  However, the statistics we have described are likely
to be features of any final formulation
of the landscape.

Indeed, these statistics appear to be robust (at least if the number
of states is finite).  They were first uncovered in studies of particular classes
of string models, but they follow from very modest assumptions:  the existence of a dense
set of states in a particular range of parameters and the absence of singularities in the
distribution of parameters, apart from those which can be understood on symmetry grounds.

Without a detailed understanding of microphysics, however, there are questions which one cannot
address.  For the landscape, the most important of these are the relative populations
of these branches, and cosmological or other effects which might select one branch over another.
At present, one can at best offer only speculative arguments why nature might find itself on one
or another of these branches:  
\begin{enumerate}
\item
The non-supersymmetric branch might be favored simply
because there might be vastly more non-supersymmetric than supersymmetric states. 
A number of constructions of such states have been exhibited\cite{silverstein,
bobkov,acharyadenef}.
Because of the need
to impose a cutoff on the supersymmetry breaking scale to control the calculation,
it is not presently possible to reliably count the states or ascertain their
statistics.  For cutoff slightly below
the fundamental scale, it seems that the numbers of supersymmetric and non-supersymmetric
states are comparable.  If there are vastly more supersymmetric than non-supersymmetric
states, than in the bulk of these states, it is unlikely that there is any fundamental
small parameter.  The supersymmetry breaking scale, internal radii, and so on are all
likely to be of order one.  In states with
small cosmological constant (in which it makes sense to
speak of a low energy theory), there
will still be a distribution of low energy parameters,
most of which, presumably, must be determined anthropically.
It is hard to understand how
such a picture can be consistent with the facts of the
Standard Model\cite{bdg}.
\item
The number of non-supersymmetric states might be highly
suppressed, as suggested by the following simple-minded stability
argument.  Supersymmetric states are stable,
so states
with very small SUSY breaking are likely to
be highly metastable.  For non-supersymmetric
states, the situation is potentially quite different.
The arguments of \cite{Denef:2004cf,dos} establish criteria for the
counting of local
minima, but global questions are more challenging.
The typical non-supersymmetric state with
small, positive cosmological constant is surrounded (in the lattice of fluxes) by other states,
presumably roughly half of which typically have negative cosmological constant.
The standard Coleman-DeLuccia analysis\cite{Coleman:1980aw} would lead
us to expect that to some of these
states the decay amplitude vanishes, while to others there is a decay to an open
universe which experiences a big crunch.  In the absence of small parameters,
one might guess that a typical state with small cosmological constant
and badly broken symmetry could decay to 50\% of its neighbors.  So if our would-be
De Sitter vacuum has, say, $300$ neighbors, then the probability that it does not undergo
rapid decay -- that it can even be thought of as a state -- is of order $(1/2)^{300}$.
It is, of course, possible that there are,
in the landscape, so many more non-supersymmetric
stationary points than supersymmetric ones that even this suppression is not important. 
It is also possible that these sorts of naive field theory ideas are not
relevant\cite{landskepticism}.
\item
The discussion
above raises the possibility
that the number of supersymmetric states might not be so much smaller than the number of
non-supersymmetric states.  In this case, due to
the logarithmic distribution of scales, a low scale of supersymmetry breaking is
reasonably probable.  The value of the weak scale might then
be a plausible accident.
{\it This is just the usual
argument for naturalness of low energy supersymmetry breaking.}    There could
well also be selection effects which prefer a small weak (and therefore
susy) breaking scale.  If there are comparable numbers of states on the
supersymmetric and non-supersymmetric branches, then there are vastly more states
on the former with a small value of the weak scale -- again, this is a version
of the usual naturalness problem.
Even if there are far more non-supersymmetric
than supersymmetric states,
it is conceivable that inflation and dark matter might might favor the supersymmetric
branch (we will discuss these possibilities
at greater length
below).  A few phenomenological facts (coupling unification, for example) provide
at least a hint
that this branch may be favored.
\item  States on the low energy branch arise when there is a (dynamically broken) R-symmetry
and supersymmetry.
These states are inevitably less numerous than states without such symmetries\cite{dgt,
dewolfeetal,dos},
and recent studies have shown that this suppression
can be quite substantial\cite{dewolfe,ds}.
One other troubling feature of this branch is that the vast majority of states
have moduli, which are only fixed as a result of supersymmetry breaking.  If the scale
of supersymmetry breaking is low, as suggested by the distribution above, then these
moduli lead to significant cosmological problems with no obvious resolution.  Note the 
$R$ symmetries under consideration here are symmetries under which the superpotential
transforms non-trivially; conventional $R$-parity is not in this category.  The analysis
of \cite{ds} indicates that states with $Z_2$ $R$-parity like symmetries are
not highly suppressed.  These would
be compatible with understanding the absence of proton decay in the intermediate branch.
\end{enumerate}

It appears difficult to fully resolve these questions in the near future.  The non-supersymmetric
branch is particularly problematic, since the supersymmetry breaking being large in virtually
all of the states, there is no small parameter which permits even the most primitive
statistical analysis.  The second and third branch are at least somewhat more accessible
to analysis.  But for the moment if one is to do any sort of ``landscape phenomenology"
one must adopt a hypothesis as to which branch of the landscape nature chooses (if any),
and  see if there is enough one can establish about the statistics of that branch
to make definite predictions.

Despite these cautionary notes,
as we have indicated, there are two arguments which might
favor the intermediate branch (the problem of metastability on the supersymmetric branch
and the suppression of the population of the third branch).
In this paper, we will adopt the hypothesis that nature lies on this branch.
We will argue that with some plausible assumptions -- assumptions
which one should be able to verify or disprove -- this branch makes definite
predictions for the spectra of gauginos and squarks and sleptons.
The assumptions involve the detailed mechanisms by which the moduli
are fixed and by which supersymmetry is broken, as well as the question
of whether matter fields live on three branes or seven branes (in the IIB
framework).
The analysis of KKLT suggests two likely possibilities for the fate of the
moduli.  In both, all moduli have masses well
above the scale of supersymmetry breaking.
In the scenario actually put forward by KKLT, there are some moduli which are
very light
compared to the fundamental scale, but heavy compared to the supersymmetry
breaking scale.  We work out the scales in some detail, finding that
the moduli can easily be several orders of magnitude more massive
than the gravitino.  In the second,
all moduli have masses of order the fundamental scale.  
We consider two possible mechanisms for supersymmetry breaking which have
been suggested for the landscape,
and argue that they are likely to lead to similar spectra for
the partners of ordinary fields.  One is the anti-brane
picture of \cite{landscape3}, developed further particularly
in \cite{nilles1,nilles2}; the second is dynamical supersymmetry breaking
in a hidden sector\cite{bdg,dgt}.   We argue that the qualitative
features of the second are identical to those of the first,
and that it is much easier to understand the nature
of supersymmetry breaking in the latter case.  We will see
that gauginos typically are significantly
lighter than squarks and sleptons.  The precise hierarchy depends on
whether there are light moduli and whether matter resides
on three-branes or seven-branes.

As we will explain,
with further work on the statistics
of gauge groups in the landscape, one can hope to verify or refute these hypotheses.
We will discuss briefly some ongoing efforts to address these issues.

Given that this branch predicts that gauginos are lighter than
squarks and sleptons, it is natural to ask why, in nature, the scalars
don't have mass approximately equal to the weak scale, while the gauginos have much
smaller mass.  Given the present limits on gaugino and higgs masses, the situation
appears quite finely tuned.  In the framework of the landscape, it is necessary to
argue that there is some selection for large gaugino masses.  The most plausible
selection criterion is a presumed requirement of cold dark matter.  If there is
an $R$ parity (likely necessary to suppress proton decay) then the lightest of the
gauginos is a natural dark matter candidate.  It might be produced thermally
or through decays of moduli or gravitinos\cite{yamaguchi,randallmoroi,kitano}.
Requiring that the dark matter be
in the correct range makes interesting predictions for the gaugino and scalar masses.

In the following sections, we develop the phenomenology and cosmology of the intermediate
scale branch.  The next section discusses supersymmetry breaking in the KKLT picture,
noting that susy breaking in hidden sectors is qualitatively similar to breaking
by $\overline{D3}$ branes.  Section 3 outlines the possible  phenomenologies, and
the resulting hierarchies of masses.   Section 4 explains how selection for the dark
matter density could account for the apparent tuning of the Higgs mass in supersymmetric
theories.  Section 5 considers both predictions of this picture, and ways it might fail.

\section{The KKLT Picture}

In their original paper, KKLT proposed a picture in which the moduli are fixed
and supersymmetry is broken, possibly with positive cosmological constant.
We will, following KKLT, consider orientifolds
of IIB theories on Calabi-Yau
spaces.  It is helpful to enumerate the basic features of the KKLT analysis;
later we will discuss which of these features are expected
to be general.
\begin{enumerate}
\item
There are two sets of moduli, complex structure moduli, $z_i$, and Kahler moduli,
$\rho$.  Choosing a set of fluxes fixes the complex structure moduli, which obtain
large masses.  The effective action for the Kahler moduli includes a superpotential,
which for large $\rho$ has the form
\beq
W = W_o +  a e^{ic\rho}.
\label{wrho}
\eeq
$W_o$ has a distribution which is flat for small $W_o$.  This fact has been verified
in explicit studies\cite{Denef:2004ze,dos},
but it is not surprising:  provided there is a dense set of states
at small $W_o$, and provided that there is nothing special about $W_o=0$,
the distribution function is non-singular and has a Taylor series expansion\cite{dos}.
For small $W_o$, the potential has an AdS stationary point at large $\rho$,
with unbroken supersymmetry.
\item  Supersymmetry can be broken if $\overline{D3}$ branes are present.
This breaking will be small if the brane is located near a warped throat.
It gives rise to an additional term in the potential for $\rho$,
\beq
V(\rho) =m_{warp}^4 {1 \over \rho^3}.
\eeq
$m_{warp}^4$ is the $\overline{D3}$ tension; it is small
due to the warping in the throat.
The distribution of $m_{warp}^2$ is also known (more precisely the distribution
of warp scales is known\cite{Denef:2004ze}:
\beq
\int dm_{warp}P_{warp}(m_{warp})=\int{dm_{warp}^2 \over m_{warp}^2 \ln(m_{warp})}
\eeq
For some fraction of states, this leads to a cosmological constant which is small
and positive.  
There have been a number of comments in the literature on the nature
of supersymmetry breaking in the anti-brane picture, and there has been
a good deal of confusion about whether the breaking should be viewed
as spontaneous or explicit. For example, it has been
suggested that this should be thought of as Fayet-Iliopoulos
mechanism, but this is problematic.
It is likely that the complications arise not only
because the models are not particularly
explicit, but because, in the warped geometry, one cannot simply write an
effective lagrangian, near $m_{warp}$, including only a finite number
of states.
\item
These last objections can be addressed by considering, instead, states in
the landscape in which
supersymmetry breaking is due to the low energy dynamics of an additional
gauge sector\cite{bdg,dgt,toappear}.  Given the assumption that chiral theories are common in the landscape,
theories with such sectors are likely to exist and are probably numerous.
As a model, one can consider the $(3,2)$ theory as a hidden sector.
There is no problem in such
a situation understanding the effective lagrangian.  The longitudinal mode
of the gravitino arises from the hidden sector.  If one writes an effective lagrangian,
the SUSY breaking
in the visible sector appears explicit.  From a phenomenological point
of view, such models are essentially identical to the anti-brane theories.  The distribution
of supersymmetry-breaking scales is the same.
The distribution of
gauge couplings is roughly flat in $g^2$\cite{Denef:2004ze}.  This corresponds to
a distribution of supersymmetry breaking scales,
\beq
\int dm_{3/2} P(m_{m_{3/2}})=\int {dm_{3/2}^2 \over m_{3/2}^2 \ln(m_{3/2})}
\eeq
just as from branes at warped throats.  
If this hidden sector lies on a $D3$ brane,
the potential behaves as $1/\rho^3$, just as in the anti-brane picture.  All of this
is consistent with the notion that in some sense these two pictures are dual
to each other.  Many other
features of the theory are easily understood in this picture.  For example, stationary
points with small cosmological constant are automatically local minima of the potential
for $\rho$.  This is because, if $\rho$ is large, 
\beq
{\partial^2 V \over \partial \rho^2} \approx e^K \vert {\partial^2 W\over \partial
\rho^2}\vert^2 g^{\rho \bar \rho}.
\eeq
The hierarchies of masses
depend on whether hidden and visible sector fields lie on $D3$ or $D7$ branes,
and are readily determined in each case.
We will enumerate the various possibilities in the
next section. 
\end{enumerate}

The KKLT analysis applies to theories with large $\rho$.  In this
case, there is, in some sense, a small parameter\footnote{The qualifier ``some
sense" is necessary because, in the IIB case, $\rho$ cannot be made arbitrarily
large.}.
But we should note that the states with small $\rho$
are also potentially important, even if they may be more difficult to
study\footnote{We
thank Shamit Kachru for stressing this point to us.}.
One expects that there are many such states, associated either with large $W_o$
or with higher order terms in the superpotential.  For these states, all of the
moduli are typically quite heavy.  Consider, first, supersymmetric states.  If the
cosmological constant is large, there is no low energy theory to discuss.  If it is small,
one can integrate out the massive fields to obtain an approximately
Lorentz-invariant, supersymmetric lagrangian for the remaining light
fields (at least the graviton and gravitino).   The low energy theory will be characterized
by a superpotential, $W_o$, and other parameters.  By our earlier arguments, the distribution
of $W_o$ (now obtained after integrating out all of the moduli) will still likely
be nearly uniform. At small $\rho$, the geometric picture is not valid.  But the
statistics of supersymmetry breaking in the DSB picture are expected to remain
the same; they rely only on some very weak assumptions about distributions
of low energy couplings.  So quite generally, we might expect the same
distribution of $m_{3/2}$ to hold for both large and small $\rho$.

We will see that, in the large $\rho$ case, $1/\rho$ serves as a small parameter.
Numerically, it's value depends on dynamical details, but if the supersymmetry is
broken within a decade or two of the TeV scale, we expect
\beq
\rho \sim - a\ln(m_{3/2}/M_p) \approx a \times 35 
\eeq
where $a$ is a number of order unity.   It is worth keeping this in mind in our subsequent
discussion.

\section{KKLT At Low Energies}

In this section, we focus on models with supersymmetry broken dynamically
in a hidden sector.  We will assume that all of the complex structure moduli
are very massive.  If the Kahler moduli are light, the low energy theory is described by $W_o$,
a superpotential for $\rho$, and an action for
the gauge and matter fields.  If the Kahler moduli are massive, there is just $W_o$.

\subsection{Structure of Soft Breakings}

As a model, we will consider,
as suggested above, a theory, such as the $(3,2)$ theory, coupled to
supergravity.
In \cite{ads}, it was argued that in models with supersymmetry dynamically broken
in a hidden sector, gauginos would be lighter than scalars. 
The argument was simple.  Suppose that $W_o$
is adjusted to (nearly) cancel the cosmological constant,
and that the scale of supersymmetry breaking is of order $\Lambda_{hid}=m_{3/2} M_p$.
In dynamical models, this is also the scale of the dynamics of the hidden
sector.  Theses models typically have no flat directions, so expectation values
of fields are of this order; $F$ terms are of order $\Lambda_{hid}^2$.  Finally,
models of DSB typically have no gauge-singlet fields.  Then,
calling the visible matter sector fields $\phi_i$, and the $(3,2)$ sector
fields $Z_i$,
there are contributions to visible sector scalar masses from a variety of
sources.
These include the terms $\vert {\partial K \over \partial \phi} W_o \vert^2$.
Terms in the Kahler potential of the form $\phi^\dagger \phi Z^\dagger Z$ will
also contribute to scalar masses.
Gaugino masses, however, are more problematic.  In the absence of singlets, one cannot
write holomorphic gauge couplings which give rise to such masses.  Anomaly mediation,
then, would typically be the leading contribution\cite{macintire,outof,lutyetal}.  

This argument is correct for the case where there are no light moduli.
However, if there are light moduli, these can acquire large $F$ terms, even in the absence of direct
renormalizable couplings to the hidden sector.  If such fields (denote them by ${\cal M}$),
have Planck scale variation, then couplings such as 
\beq
K_{{\cal M} \phi}={\cal M}^\dagger {\cal M} Z^\dagger Z
+ {\cal M} Z^\dagger Z + \dots 
\eeq
will generate order one shifts in ${\cal M}$, and order $m_{3/2} M_p$ values for the
Kahler derivative ${\partial K \over \partial {\cal M}} W.$
So light moduli, in the presence of dynamical supersymmetry breaking, can generate
${\cal O} (m_{3/2})$ masses for gauginos.  Light moduli can have other effects as well.

In the IIB flux vacua, we have enumerated two
possibilities.  First, there may be no light moduli.  In that
case, the arguments of \cite{ads} would be correct, and gauginos would typically be light.
Since the $W_o$ distribution is the
same in both the light and heavy
Kahler moduli cases, there is no reason to think that there are significantly more states
of one type or the other.   It is thus a phenomenological
question whether one is on one branch or the
other.  One can advance some phenomenological arguments
against the heavy moduli branch.
\begin{enumerate}
\item  While gaugino masses may be lighter than squark and slepton masses, flavor changing
neutral currents are not likely to be suppressed.  (Note that there is no obvious anthropic 
argument which might select for, say, squark or slepton degeneracy).
\item  Given that there is no apparent small parameter, it is not clear how coupling
unification might emerge (though we do not have a compelling picture for how unification
will emerge in the light modulus, large $\rho$, case, the existence of a small
parameter which controls the coupling is promising.).
\end{enumerate}

\subsection{Light Kahler Moduli; Hidden Sector on $D3$ Branes}

In the case where some of the Kahler moduli are light, a more detailed analysis is required.
The form of the spectrum depends on whether the hidden sector and matter fields live
on $D3$ or $D7$ branes.
For simplicity, consider the case of a single complex structure modulus (the result generalizes to several
moduli).  If the hidden sector chiral fields, denoted by $z_i$,
reside on $D3$ branes, for large $\rho$ the Kahler potential is
\beq
K = - 3 \ln(i(\rho -\rho^\dagger) - z_i^\dagger z_i).
\eeq
The gauge couplings in the hidden sector -- and thus the scale of supersymmetry
breaking, $\Lambda_{hid}$, -- are independent of $\rho$.
Suppose, as in the KKLT picture, one has found the solution of the condition
\beq
D_\rho W =0.
\eeq
For small $W_o$ in eqn. [\ref{wrho}],
$\rho$ is large, $\rho \sim -\ln(W_o)$.
Now imagine ``turning on" the hidden sector.  This generates a term in the energy
$\Lambda_{hid}^4$.  For states with small cosmological constant, this is equal
to $3 \vert W_o \vert^2/\rho^2$.  As a result, there is a supersymmetry-violating potential for the
Kahler modulus,
\beq
V(\rho) \approx {1 \over \rho^3} (3\vert W_o \vert^2)
\eeq
$$~~~~~\approx {m_{3/2}^2}.$$

We first estimate the mass of $\rho$, then the $\rho$ tadpole
and $F$ term.  The largest contribution comes from the
supersymmetric part of the potential, eqn. [\ref{wrho}].  Because $\rho$ is large,
we need only differentiate the exponential terms.  Also, because (in the absence
of the hidden sector) supersymmetry is unbroken, we have
\beq
{\partial W \over \partial \rho} \approx {3 \over \rho} W_o.
\eeq
So the mass-squared of the $\rho$ field (recall the $\rho$ kinetic term is proportional to
$1/\rho^2$ is positive and hierarchically large compared to the gravitino mass:
\beq
m_{\rho}^2 \approx \rho^2 m_{3/2}^2.
\eeq
In particular, it is large compared to $m_{3/2}$; the $\rho$ spectrum is approximately 
supersymmetric.

Treating $V_{hid}$ as a perturbation, we can determine the shift in $\rho$:
\beq
\Delta \rho = {\partial V_{hid} \over \partial \rho}/m_{\rho}^2
\eeq
$$~~~~~\approx {1 \over \rho}.$$
Finally, the $\rho$ F term follows from the equations of motion (keeping in mind
the form of the $\rho$ kinetic term)
\beq
F_\rho \approx W_o.
\eeq

Now we can ask the form of the soft breakings induced for visible sector fields.
These depend, again, on whether the matter fields (chiral fields and
gauge fields) reside on $D3$ or $D7$ branes.  In the case of $D3$ branes,
gaugino masses receive no contributions from $F_\rho$, and the leading contributions
to the gaugino masses are the anomaly mediated ones.  The Kahler
potential for the matter fields has a
no-scale form.  So the leading, ${\cal O}(m_{3/2}^2)$ terms vanish.   
The $F_{\rho}$ contribution to the masses is of order
\beq
m_\phi^2 \approx {1 \over \rho^2} m_{3/2}^2.
\eeq
If $\rho$ is of the order we have estimated
earlier, then this is comparable to, and
possibly somewhat larger than, the anomaly-mediated contribution to the
squark and slepton masses.  There can be additional contributions, arising from
$\alpha^\prime$ corrections to the Kahler potential, as well as one loop effects
in the low energy theory.  The $\alpha^\prime$ corrections have been computed\cite{hl}, and
are of order $1/(\rho)^{3/2}$.  This correction has been estimated as between $10\%$
and $1\%$\cite{largevolume,fixingallmoduli},
corresponding to scalar masses of order $1/3-1/10$ of the gravitino mass.

If the Standard Model fields live on $D7$ branes, the Kahler potential no longer
has the sequestered form, and the squark and slepton masses are simply of order $m_{3/2}$.
Gaugino masses are still suppressed, though now the coupling of $\rho$ to the
gauge fields leads to a contribution of order $1/\rho$ to gaugino masses beyond
just the anomaly mediated contribution. 
 
\subsection{Hidden Sector on $D7$ Branes}

If the hidden sector lies on $D7$ branes, the scalings are different.  Now $\Lambda_{hid} $
depends on $\rho$, $\Lambda_{hid} \propto
e^{a\rho}$.  But the relation between $\Lambda_{hid}$ and $W_o$ is also
altered, because of the different structure of the Kahler potential:
\beq
\Lambda_{hid}^4 = 3\vert W_o\vert^2.
\eeq
So $\delta \rho \sim 1.  $
and
the supersymmetry breaking mass term for $\rho$ is comparable in size
to the non-supersymmetric term: 
\beq
\delta m_\rho^2= \rho^2 m_{3/2}^2. 
\eeq
Correspondingly, $F_\rho$ is large:
\beq
F_\rho \approx \rho m_{3/2}.
\eeq

Again, the masses of the visible sector fields depend on whether they reside on $3$
branes of seven-branes.  In both cases, the squark and slepton masses are of order 
$m_{3/2}^2$.  If the gauge fields reside on $7$ branes, because $\rho$ couples
directly to the gauge fields and $F_\rho$ is large, the gaugino masses are of order
$m_{3/2}$.  If they reside on $D3$ branes, again the principle contributions are the
anomaly mediated ones.

\section{Phenomenology and Cosmology}

In the landscape, one would hope, on some principled grounds, to argue that
some class of states are selected, and that this would lead to some definite phenomenological
predictions.  Such selection is likely to involve a combination of
anthropic considerations, statistical features (e.g. far more of one type of state
than another), and cosmological considerations.
As the discussion of low energy supersymmetry in the introduction indicates,
our current understanding is too primitive to make firm statements at present.
But we can at least delineate some of the basic issues and speculate as to how
predictions may emerge.  In the case of low energy supersymmetry, we gave some arguments
why supersymmetric states might be favored, and discussed expectations for the
features of the statistics of supersymmetry breaking which follow from known features
of the landscape.  At a more detailed
level, we advanced an argument that discrete $R$ symmetries (apart from $Z_2$ symmetries)
are quite uncommon.  Based on these considerations, we focussed most of our energy
on the features of the intermediate scale branch.

Within this branch we have seen there are still a variety of choices:  supersymmetry
breaking could be due principally to fields on $D3$ or $D7$ branes, and the
fields of the Standard Model could reside on one sort of brane or the other.  We have
also seen that there may be light moduli, or there may not.
It might well be possible to determine that one or
another of these phenomena is common or rare, just as we have argued that
discrete $R$ symmetries are rare.  For the moment, we can simply note that some
of these possibilities are more readily reconciled with facts of nature than others.

There is an important question about which the landscape might shed some light.
Low energy supersymmetry, in the post LEP II era, has become less
attractive as an explanation of the hierarchy problem.  Elaborate constructions
are necessary to avoid tuning at the few per cent level.
In the landscape framework,
one would like to find a selection principle which accounts
for this tuning.  Perhaps the most plausible candidate
is the density of dark matter (in combination with a selection
for the scale of weak interactions).  The dark matter density is closely tied, in modern
theories of structure formation, to the features of galaxies.  It has been suggested
from time to time that anthropic considerations might fix the dark matter density.
There is currently no convincing calculation which demonstrates this,
but it is conceivable that this is the
case, and, if so, we can ask where this might lead.  In many of the models
we have been studying in this paper, if the gaugino masses are fixed
by dark matter considerations, to be in the several hundred GeV range, the squark and
slepton masses will be $10$'s of TeV.  The Higgs mass in a typical vacuum will
also be of this order; only in a small fraction will the $W$ and $Z$ bosons be light.
This, however, might be selected by other considerations, which have been widely
discussed in the literature.  In the rest of this section, we show that selection
for the observed dark matter density and the weak scale leads to a cosmology
for moduli and gravitinos which is compatible with known observations, as well
as a predictive supersymmetry phenomenology, compatible with known facts.
 
\subsection{Cosmology}  

We will focus in this section on the cosmology of the large $\rho$ (light $\rho$) case.
Usually, the cosmology of moduli with masses comparable to the weak scale is problematic.
These moduli come to dominate the universe long before recombination, and their decays
tend to destroy the light elements produced during nucleosynthesis.  One suggestion
to resolve this dilemna\cite{bkn} is to suppose that the moduli are relatively heavy.  A $10$ TeV
modulus might be expected to reheat the universe to about $10$ Mev, restarting nucleosynthesis.
Normally, this is said to involve an unnatural fine tuning.  But we have just seen that
in the KKLT model, $\rho$ is much heavier than the squarks and sleptons.
We can easily imagine, in fact, that $m_{\rho}$ is of order $1000$ TeV or
even larger!  We have seen
that this is not fine tuned, because the bulk of the $\rho$ mass is {\it supersymmetric}.

It is necessary that one produce baryons quite late in such a cosmology, but
this could occur through A-D baryogenesis\cite{adbaryogenesis}.
More serious are the questions
of producing adequate dark matter and avoiding overproduction
of gravitinos.  In a similar framework, these issues have been dealt
with in \cite{yamaguchi}.      
The reheat temperature at $\rho$ decay is approximately:
\beq
T_{rh} = m_{\rho}\left ( {m_\rho \over M_p} \right )^{1/2}
\eeq
$$~~~~~\approx 0.2 {\rm GeV} \left ( {m_{3/2} \over 10 {\rm TeV}} \right )^{3/2}
(\rho/35)^{3/2}.$$
In the decays of $\rho$, direct production of both gauginos
and gravitinos is suppressed by chirality. 
The fraction of the energy density initially in gravitinos
is of order:
\beq
{\rho_{3/2} \over \rho_{total}} = C {m_{3/2}^3 \over m_{\rho}^3}\sim \rho^{-3}.
\eeq
Here we have adopted the notation of \cite{yamaguchi}; $C$ is plausibly of order $0.1$-$0.01$.
The gravitino lifetime,
on the other hand, is $\rho^3$ longer than the modulus lifetime, corresponding to
decay when the temperature has dropped by a factor $\rho^{3/2}$.  The temperature at gravitino
decay is thus of order:
\beq
T_{3/2}\approx 1 {\rm MeV} \left ( {m_{3/2} \over 10 {\rm TeV}} \right )^{3/2}.
\eeq
Even though the gravitinos still constitute a small fraction
of the energy density (of order $C \rho^{-3/2}$, one has to worry that gravitino decay products
will destroy $De$, $Li$, and other light elements\cite{moroi}.
However, as discussed at length in \cite{yamaguchi}, there are a plausible range
of parameters for which this is not a problem.

Our real question, however, is the dark matter density. As stressed in \cite{yamaguchi},
gravitino decays will produce of order one LSP per decay.  Suppose that the mass of the 
LSP is given by the anomaly mediated formula; then, one expects
\beq
m_{LSP} \sim 3 \times 10^{-3} ~m_{3/2}.
\eeq
So the fractional density at $1$ eV is:
\beq
{\rho_{lsp} \over \rho_{\gamma}} = 2 \times 10^{-1} (C/10^{-2})  
\left ( {m_{3/2} \over 10 {\rm Tev}} \right )^{3/2} ({\rho \over 35})^{-3/2}.
 \eeq
 So we see that we account for the dark matter density with an LSP
 which is in the right range to have escaped detection in accelerators, while
 the gravitino -- and the scalar masses -- are in the TeV range or higher.
 
 \subsection{Phenomenology}
 
 We will leave a thorough study of the phenomenology of these theories to further work.
 The basic point is that we have seen that selection
 for the dark matter density and the weak scale could
 account for the high masses of the
 squarks and sleptons, while implying that gauginos should
 be produced at the LHC.  So the fact that LEP II did not discover
 supersymmetry is not surprising from this viewpoint.

\section{Conclusions:  Selection effects, the little hierarchy, and other issues}

While low energy supersymmetry has many attractive features, experiments have
narrowed significantly the possible parameter spaces of supersymmetric models,
and it is generally believed that the lightness of the Higgs particle in supersymmetry
constitutes a significant fine tuning problem.
The cosmological argument we have presented above provides a possible
resolution of this puzzle, provided we are willing to invoke a weak anthropic
argument.  The existence of structure probably requires the existence
of cold dark matter, with a density (all other constants of nature held fixed) in
a limited range.  If the density is too low, fluctuations do not grow large until
the universe is dark energy dominated.   If the density is too high, it is possible
(though by no means certain)
that features of the resulting structure are inhospitable\cite{aguirre}.  Limits of the
latter sort are subjects of debate; here we have simply
assumed that such
considerations limit the dark matter density to a narrow range.  This in turn limits
the lightest gaugino (the LSP in our scenario) to a narrow range of mass; squarks and
sleptons are generically one or two orders of magnitude
more massive.  If we assume that the weak scale must also lie within
a narrow range, then there would be a selection for the apparently tuned set of parameters
which this model requires.

As usual, in any such discussion of anthropic selection, we cannot establish with any
certainty that variation of several parameters -- the gaugino masses, the inflationary
fluctuation spectrum, the weak scale, and so on, one can't find other points consistent
with the existence of observers, or whatever one feels is the correct
selection criterion.   We are here
adopting the point of view that there {\it may} be selection for any parameter a drastic
change of which would be devastating to the existence of life as we know it.  We are establishing,
at best, that it is plausible that a particular set of states in the landscape is preferred;
we can then ask -- as we have here -- what are the consequences for experiments
of such a preference.

With the assumptions we have made here, we see that a number of statements are robust.
It is likely that in an order one fraction of states,
squarks and sleptons are generally significantly more massive than gauginos.  The moduli
are quite heavy -- either heavy enough that they restart nucleosynthesis, or so heavy
that they play no role at all in low energy physics.  In the former case, the squarks and
sleptons are expected to be an order of magnitude or so more massive than the gauginos;
in the latter, the gaugino masses arise from anomaly mediation, and the masses are separated
by a full loop factor.

The resolution of the cosmological moduli problem, and the possible explanation of
supersymmetric fine tuning are two attractive features of the picture we have
developed.  There are still many questions.  
\begin{enumerate}
\item  While squarks and sleptons can be significantly heavier than usually assumed, they are
not heavy enough to resolve the flavor problems of supersymmetric theories.  Some approximate
degeneracy or alignment is still required.  In the landscape, one must argue that this is
either typical or that there is some effect which selects for such symmetries.  One possibility
is that there is some large class of Kahler potentials for which, at lowest order in $\rho$,
squarks and sleptons are degenerate.  The combination of somewhat heavier scalars and
approximate degeneracy could resolve some of the questions of flavor.
\item  Strong CP is a puzzle.  There is no obvious, generic
light axion candidate; because the $\rho$ field is approximately
supersymmetric, there is no light pseudoscalar
here.  We are assuming that the lsp is the dark matter particle, so
selection for dark matter is not likely to produce an axion.  
\item  Inflation and the Brustein-Steinhardt problem: it would be appealing if the
field $\rho$ could somehow play the role of the inflaton.  This may be possible, but it does
not follow in an obvious way from the features of the $\rho$ dynamics which we have outlined.
If a single Kahler modulus were to play the role of inflaton, then with the scales
we have assumed here, the quantum fluctuations would be too small to account
for structure.  Multiple Kahler moduli might lead to hybrid inflation, but further
tuning of potential parameters, at at least the $1\%$ level, would be required
to obtain adequate inflation and fluctuations.  The $\rho$ fields also potentially suffer
from the Brustein-Steinhardt problem.  This might be solved by features of
some early, high energy, period of inflation, along lines suggested in \cite{dinebs}.
If some other field is responsible for inflation and dominates the energy for
a time, the potential for $\rho$ can be appreciably altered in a way which
dramatically slows the motion of the field.
\end{enumerate}

There are many ways, as we have indicated, in which the ideas described here might 
fail.  Perhaps the most dramatic is that the landscape
may not exist\cite{landskepticism}, or alternatively that there might
exist infinite numbers of states\cite{infinite} whose existence
might require significant rethinking of our basic understanding
of string theory and what it might have to do with nature.
But we believe we have outlined plausible predictions of a broad swath of states
within the landscape.  Further work could establish, or disprove, these ideas.
It is important to do a more thorough analysis
of the cosmology of these models as a function of the parameters $\rho$
and $m_{3/2}$, and this is in progress.  
One area for further study is the problem of dynamical supersymmetry
breaking within the landscape.
This is related to the problem of understanding distributions of gauge groups and
matter content.  Underpinning the structure we have studied in this paper
is an assumption that hidden sectors without gauge
singlets are generic, but this seems a question that one should be able to answer.

\noindent
{\bf Acknowledgements:}
We thank A. Aguirre, N. Arkani-Hamed,
T. Banks, A. Birkedal, S. Kachru, E. Silverstein and S. Thomas for conversations.
This work supported in part by the U.S.
Department of Energy.

\end{document}